\documentclass[aps,pre,twocolumn,preprintnumbers,amsmath,amssymb,nofootinbib]{revtex4-2}
\usepackage{epsfig}
\usepackage{epstopdf}
\usepackage{lipsum}

\begin{document}
	\title{Scientific success from the perspective of the strength of weak ties}

	\author{Agata Fronczak, Maciej Mrowiński and Piotr Fronczak}
	\affiliation{Faculty of Physics, Warsaw University of Technology,
		Koszykowa 75, PL-00-662 Warsaw, Poland}
	\date{\today}
	
	\begin{abstract}
We present the first complete confirmation of Granovetter's theory of social networks using a massive dataset. For this purpose, we study a scientific collaboration network, which is considered one of the most important examples that contradicts the universality of this theory. We achieve this goal by rejecting the assumption of the symmetry of social ties. Our approach is grounded in well-established heterogeneous (degree-based) mean-field theory commonly used to study dynamical processes on complex networks. Granovetter's theory is based on two hypotheses that assign different roles to interpersonal, information-carrying connections. The first hypothesis states that strong ties carrying the majority of interaction events are located mainly within densely connected groups of people. The second hypothesis maintains that these groups are connected by sparse weak ties that are of vital importance for the diffusion of information - individuals who have access to weak ties have an advantage over those who do not. Given the scientific collaboration network, with strength of directed ties measured by the asymmetric fraction of joint publications, we show that scientific success is strongly correlated with the structure of a scientist's collaboration network. First, among two scientists, with analogous achievements, the one with weaker ties tends to have the higher h-index, and second, teams connected by such ties create more valuable publications.
	\end{abstract}
	\maketitle
	
Social networks (SN), representing patterns of human interactions, have been the subject of both empirical and theoretical research since at least the middle of the last century \cite{1994bookWasserman}. At the beginning of the 21st century, there was a breakthrough in social network analysis (SNA) \cite{2012manifestoSNA, 2020ScienceLarez}. With the era of widespread digitization, which provided access to huge electronic databases, new empirical methods of SNA have emerged and replaced traditional approaches based on questionnaires and interviews. These new methods, rooted in big data mining, finally allowed for the verification of many well-established theoretical SN ideas, in some cases confirming their validity and in others failing to do so \cite{2012NatureGiles}. In this regard, the present status of Granovetter's weak-tie theory \cite{1973Granovetter, 1995bookGranovetter} of SN, one of the oldest and most influential theories in sociology, is still vague. There are convincing studies that show the validity of its selected aspects (e.g., \cite{2007PNASOnnela, 2010ScienceEagle, 2012NatPhysPajevic, 2012PLoSGrabowicz}), but there are also many that question it (e.g., \cite{2012EPLPan, 2016JAmSocAral, 2017JLaborEconGee}). Our analysis presented in this paper is unique because, using a massive dataset, not only do we confirm Granovetter's weak tie theory in its full spectrum but also indicate a possible source of problems related to research questioning its validity.

Granovetter's theory is based on two hypotheses. The first pertains to the structure of social networks and the second to their dynamics (the way in which the afore-mentioned structure influences the flow of information in the network). It is significant that although most empirical studies have focused on the first hypothesis, far less research has been undertaken to verify the second. One possible reason is that the second hypothesis involves notions relative to the nature and importance of information that are hard to quantify and measure. In this study, we clearly confirm both hypotheses - and Granovetter's theory in its entirety - in the context of a scientific collaboration network. 

The scientific collaboration network \cite{2001PNASNewman, 2001PREaNewman, 2001PREbNewman, 2002PhysABarabasi} is particularly well suited to the overarching goal of this paper (i.e., complete confirmation of Granovetter's theory) because i) connections (ties) between network nodes (scientists) are well defined, and their weight\footnote{\label{footnote1}In complex networks, the Granovetter's concept of \textit{tie strength} corresponds to \textit{edge weight}, while the concept of strength refers to the network nodes and is defined as the total weight of their connections \cite{2004PNASBarrat}. Due to historical reasons, in this paper the notions of: tie strength and edge weight are treated as equivalent and used interchangeably.} (strength of ties) is easy to measure (e.g., through joint publications); ii) scientific publications themselves are also a specific proxy of information flow in the studied network (diffusion of innovations \cite{2003bookRogers}); and iii) the number of citations is an obvious measure of their significance. Easy access to large datasets is also important, making our conclusions statistically reliable.

The network we investigated has all the features of a complex network \cite{2010bookNewman}. In particular, it shows the scale-free node degree distribution with the characteristic exponent $\gamma\simeq 2.5$. In the theory of complex networks, this value of $\gamma$ is alarming in the sense that it indicates that the network requires special treatment, including methods of results averaging different to the ones used in homogeneous systems. In relation to Granovetter's theory, this means that in such networks, basic concepts, such as tie strength and neighbourhood overlap, should be defined in a more careful manner than in homogeneous networks. Their incorrect definition may, instead of confirming the theory, result in its contradiction. In all known empirical studies on Granovetter's theory, interpersonal ties are assumed to be positive and symmetric. However, it is obvious that social relations do not usually follow this assumption. For example, the scientific collaboration between a young scientist and an established one can hardly be called symmetric. 

\begin{figure*}[t]
	\includegraphics[width=2\columnwidth]{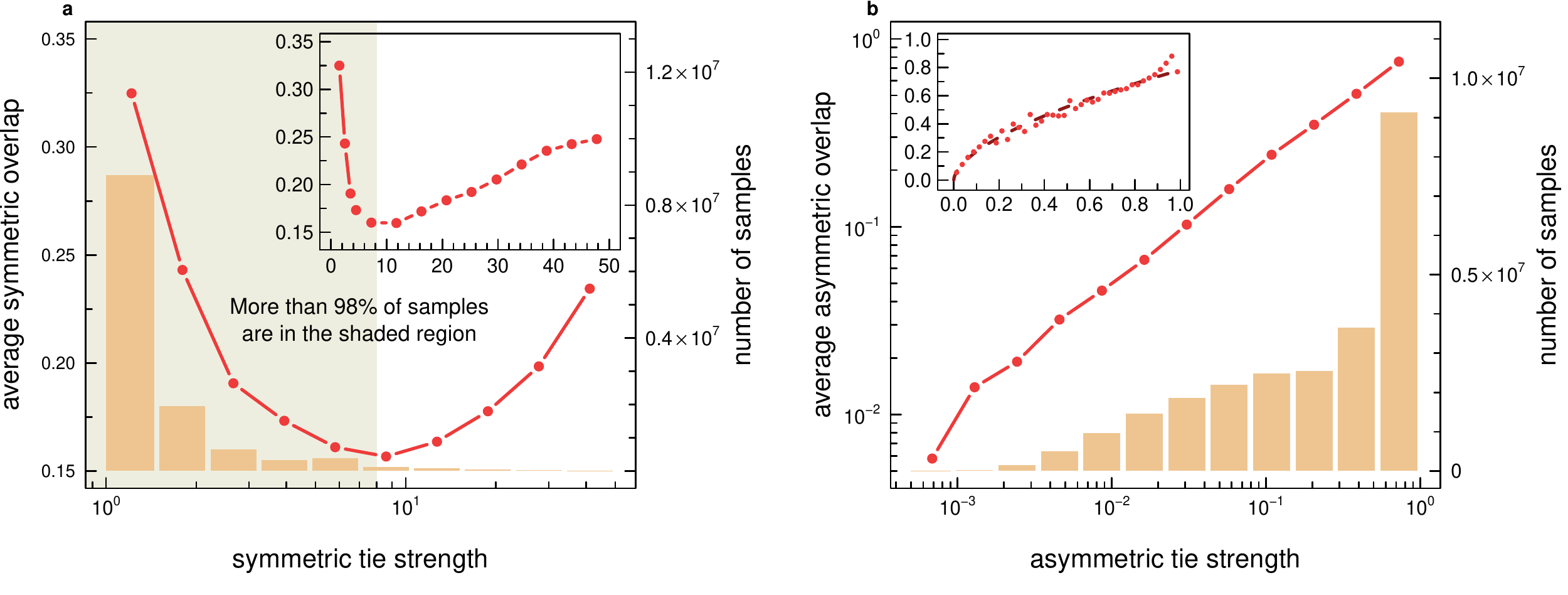}
	\caption{\label{OijQij} \textbf{Dependence of neighbourhood overlap on tie strength} in: \textbf{a} -- undirected, weighted DBLP scientific collaboration network, in which tie strength $w_{ij}$ corresponds to the number of joint publications (i.e. the number of times that co-authorship has been repeated) and the symmetric neighbourhood overlap $O_{ij}$ is given by the standard formula, Eq.~(\ref{Oij}); \textbf{b} -- directed, weighted projection of the same network with asymmetric tie strength $v_{ij}$ and asymmetric overlap $Q_{ij}$, obtained from Eqs.~(\ref{vij}) and~(\ref{Qij}). In both graphs, circles indicate averages of overlaps (in intervals of logarithmically increasing width in the main panel and of constant width in the inset, respectively), while bars represent the number of samples from which the averages were calculated. Empirical relationships, similar to the one from the left graph (\textbf{a}), showing the decreasing character of $O_{ij}(w_{ij})$, have so far been the basic argument against validity of the Granovetter's theory in scientific collaboration networks. The graph on the right (\textbf{b}) shows that the necessary condition to confirm the Granovetter's theory in the studied networks is to reject the assumption about the symmetry of social ties.}
\end{figure*}

In his original paper \cite{1973Granovetter}, Granovetter treated ties as if they were positive and symmetric, but he also noted that "the comprehensive theory might require discussion of negative and/or asymmetric ties". We follow this suggestion in this study and reject the assumption about the symmetry of social ties, which is omnipresent in the literature on the subject. The validity of this approach can be explained by intuition trained in the field of complex networks. Granovetter argued that "the degree of overlap of two individuals' friendship networks varies directly with the strength of their tie to one another". However, from the theory of complex networks, we know that in social networks with a high degree of heterogeneity (e.g., due to scale-free node degree distribution), the sizes of ego-networks of two connected nodes may differ drastically. Therefore, their common neighbours can be a significant part of the neighbourhood of one node and an insignificant part of the neighbourhood of the other, resulting in a completely different perception of the strength of the link on both ends.

In what follows, we show that the above reasoning, which assumes the asymmetry of tie strength, allows for a quantitative validation of Granovetter's theory in scientific collaboration networks, that have resisted such verification so far. We use the DBLP Computer Science Bibliography dataset, which includes information on nearly five million computer science papers (i.e., their publication dates, lists of authors and citation records) authored by over four million scientists (see \textit{Methods} for more details).

\section*{Results}

In the standard approach to scientific collaboration networks, the nodes represent authors, and an undirected internode connection occurs when two authors have published at least one paper together. When considered as binary networks - without any additional features assigned to nodes and connections - these networks show numerous structural similarities to other SNs (e.g., skewed degree distribution, high clustering and small-world effect) \cite{2001PNASNewman, 2001PREaNewman, 2001PREbNewman}. However, when edges are assigned weights representing, for example, the number of joint publications, then, although macroscopic characteristics of scientific collaboration networks (e.g., distributions of connection weights and node strengths) still correspond to those observed in typical SNs \cite{2004PNASBarrat, 2007PNASOnnela}, their microscopic structure related to the location of strong and weak ties is completely different. Dense, local neighbourhoods of nodes consist of weak ties, while strong ties act as bridges between local research groups. The atypical properties of scientific collaboration networks have been confirmed in several independent studies \cite{2012NatPhysPajevic, 2012EPLPan, 2014PREKe}. 

\begin{figure*}[t]
	\includegraphics[width=2\columnwidth]{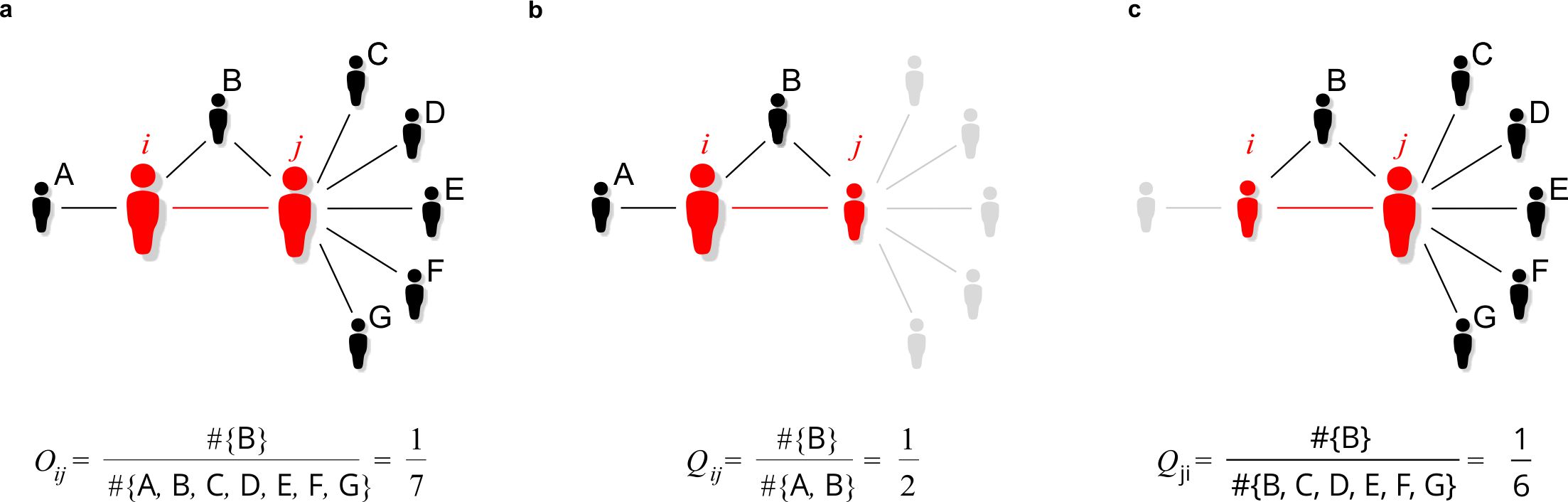}
	\caption{\label{overlap} \textbf{Illustration of the difference between symmetric and asymmetric neighbourhood overlap}. In the figure, to highlight the benefits of analysing asymmetric overlaps, the corresponding values of: \textbf{a} --  symmetric $O_{ij}$ (\ref{Oij}) and \textbf{b}, \textbf{c} -- asymmetric $Q_{ij}\neq Q_{ji}$ (\ref{Qij}) overlaps have been calculated for the same network configuration, in which interconnected nodes differ in the size of their ego-networks. In such cases, which are typical for complex networks with underlying fat-tailed distributions, a common scenario is that for $k_i\ll k_j$ one has $Q_{ij}\gg Q_{ji}\simeq O_{ij}$. This explains why introducing tie direction is necessary for reliable verification of the Granowetter's theory in scientific collaboration networks.}
\end{figure*}

Specifically, as shown in Ref.~\cite{2012EPLPan}, these unusual weight-topology correlations can be seen by analysing the relationship between the tie strength, $w_{ij}$, of two scientists $i$ and $j$, and the overlap, $O_{ij}$, of their ego-networks. As indicated by Onnela et al. \cite{2007PNASOnnela, 2007NewJPhysOnnela}, the overlap of two connected individuals is the ratio of the number of their common neighbours, $n_{ij}$, to the number of all their neighbours:
\begin{equation}\label{Oij}
O_{ij}=\frac{n_{ij}}{(k_i-1)+(k_j-1)-n_{ij}},
\end{equation} 
where $k_i$ and $k_{j}$ represent degrees of the considered individuals. In typical SNs, the above-defined overlap is an increasing function of the tie strength, $w_{ij}$,  while analyses of scientific collaboration networks show something completely different. As can be seen in Fig.~\ref{OijQij}a, in the studied network of computer scientists, with $w_{ij}$ standing for the number of joint publications\footnote{\label{footnote2}It should be noted that the number of joint publications, which corresponds to the number of times a collaboration between two scientists has been repeated, is not the only possible choice for the tie strength. For example, in Refs.~\cite{2012EPLPan, 2014PREKe, 2008PRLOpsahl} the formula introduced by Newman \cite{2001PREbNewman} is used: $w_{ij}=\sum_p\frac{1}{n_p-1}$, where $p$ is the set of papers co-authored by $n_p$ scientists, including $i$ and $j$. The motivation behind the Newman's formula is that an author divides his/her time and other resources between $n_p-1$ collaborators, and thus the strength of the connection should vary inversely with $n_p-1$. However, in comparison to the definition we use: $w_{ij}=\sum_p 1$, Newman's formula does not take into account synergy effects of working in a group, nor the effect of social inertia \cite{2006PRERamasco, 2007EPJRamasco} that measures the tendency of scientists to keep on collaborating with previous partners, which seem important in the context of scientific collaboration networks.}, for the vast majority of connections ($98\%$), the overlap decreases with connection weight. This relationship indicates that weak ties mainly reside inside dense network neighbourhoods, whereas strong ties act as connectors between them. It has been hypothesized that this counterintuitive observation could be attributed to different driving mechanisms of tie formation and reinforcement in scientific collaboration networks in comparison to other social networks \cite{2012EPLPan}. In what follows, we argue that the observation is related to the definitions of the tie strength and neighbourhood overlap that are not properly suited to the structure of the studied network.

First, let us deal with the definition of the overlap (\ref{Oij}) (referred to as \textit{symmetric overlap}). In Fig.~\ref{overlap}a, this local measure is shown in the case of a link connecting nodes with significantly different degrees. In such cases, for $k_i\ll k_j$, Eq.~(\ref{Oij}) can be simplified to $O_{ij}\simeq n_{ij}/k_j$, which shows that it is strongly biased towards nodes with high degrees, distorting the image of the common neighbourhood as seen from the perspective of nodes with small degrees. This drawback of symmetric overlap gains importance in networks with highly skewed, fat-tailed node degree distributions $P(k)$. In such networks, as brilliantly exploited by the degree-based mean-field theory of complex networks \cite{HMF1, HMF2, HMF3}, node degree distributions for nearest neighbours are even more fat-tailed than the original distributions $P(k)$. As a result, the number of edges in such networks connecting nodes with high and low degrees can be very high, leading to an unintended overrepresentation of strongly connected nodes by Eq.~(\ref{Oij}).

\begin{figure*}[t]
	\includegraphics[width=2\columnwidth]{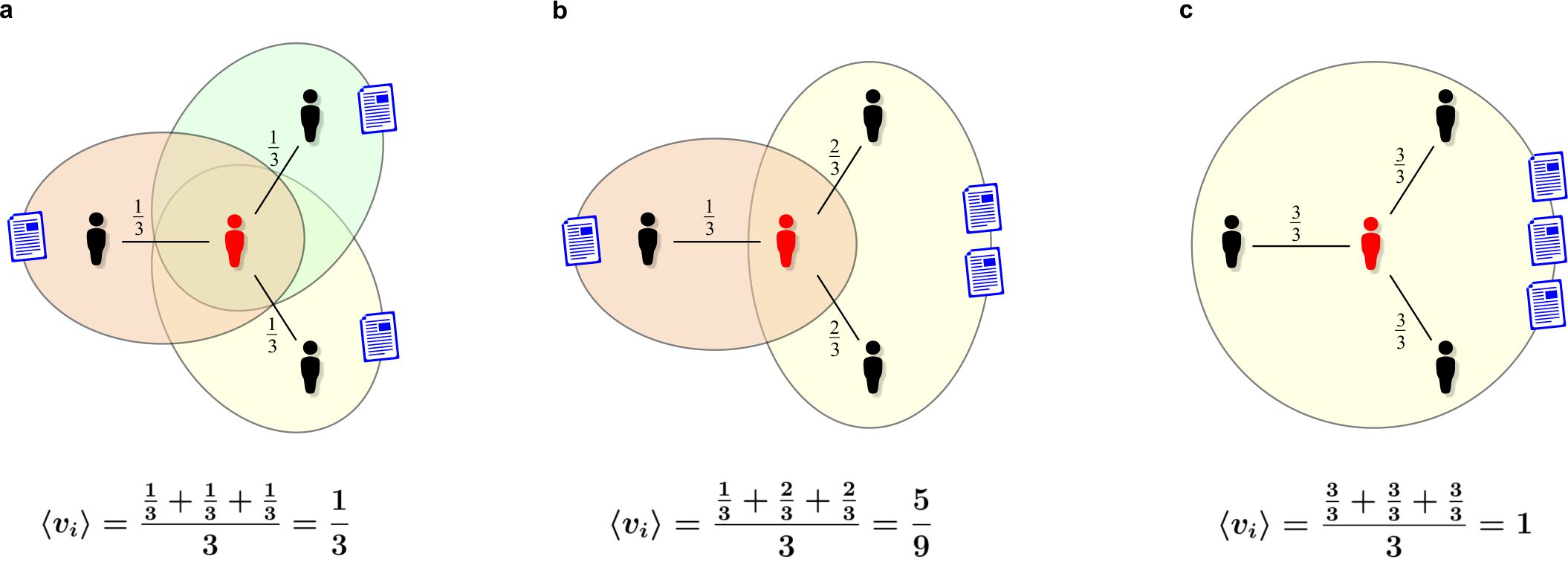}
	\caption{\label{meanv} \textbf{Average asymmetric tie strength of a scientist}. The figure presents ego-networks of three different scientists (egos) with the same number of co-authors $k_i=3$ and publications $p_i=3$, but with different patterns of collaboration. On the left scheme \textbf{a}, each of the three publications has only two authors; on the central scheme \textbf{b}, two publications were written by a team of three and one by a team of two; in the scheme on the right \textbf{c}, all publications involved the entire ego-network of a scientist. In each of the presented cases, the ego's average asymmetric tie strength is different. Its value increases from the left diagram to the right, exactly in the same way as the intuitively understood social role of collaborators, on which depends not only the ego's productivity but also integrity of his/her research group.}	
\end{figure*}

To overcome problems with symmetric overlap, we introduce the concept of \textit{asymmetric overlap}:
\begin{equation}\label{Qij}
	Q_{ij}=\frac{n_{ij}}{k_i-1}\neq Q_{ji}.
\end{equation}
This can be used to describe the overlap between the neighbourhoods of two connected nodes from the perspective of each node separately. In the context of complex networks, this new definition is free from the shortcomings of the previous one. In particular, it copes well with connected nodes (collaborating scientists) whose degrees (ego-networks) differ significantly - that is, when their common neighbours (if any) are a significant part of the neighbourhood of one node and an insignificant part of the neighbourhood of the other. In such cases, the values of $Q_{ij}$ and $Q_{ji}$ corresponding to the same tie are different (see Fig.~\ref{overlap}b,c). The value of $Q_{ij}$ that is close to 1 means that almost all neighbours of $i$ are also neighbours of $j$. The value of $Q_{ji}$ close to 0 means that only a small part of the neighbourhood of $j$ belongs to the neighbourhood of $i$.

The concept of asymmetric overlap naturally leads to the idea of directed networks and justifies the introduction of \textit{asymmetric tie strength}:
\begin{equation}\label{vij}
	v_{ij}=\frac{w_{ij}}{p_{i}}\neq v_{ji},
\end{equation} 
where $p_i$ stands for the number of all publications of the $i$-th scientist\footnote{\label{footnote3}Note that the number of publications does not have to be equal to the strength of the node: $p_i\neq s_i=\sum_j w_{ij}$. It results from the definition of symmetric tie strength $w_{ij}$ adopted in this publication, which we commented on in the footnote~\ref{footnote2}.}. The intuitive rationale behind Eq.~(\ref{vij}) is as follows: For a young scientist, with a small number of publications, each publication makes a significant contribution to his or her publication output, just as each co-author is an important part of his or her research environment (cf. Eqs.~(\ref{Qij}) and~(\ref{vij})). However, the importance of each publication and collaboration from the perspective of an established scientist with a large number of publications and an extensive network of collaborators is completely different. Depending on the circumstances, a given number of joint publications (e.g., $w_{ij}=1$) may have a completely different meaning.

In Fig.~\ref{OijQij}b, the dependence of asymmetric overlap on asymmetric tie strength for the considered network of computer scientists is shown. Contrary to what can be seen in Fig.~\ref{OijQij}a, the relationship $Q_{ij}(v_{ij})$ is increasing in the entire range of variability of its parameters. The result indicates that, from the point of view of a single scientist (ego-network approach), strong ties mainly constitute dense local clusters, whereas weak ties connect these clusters or play the role of intermediary ties \cite{2012PLoSGrabowicz}. The observation clearly confirms the validity of Granovetter's first hypothesis in scientific collaboration networks. 

\begin{figure*}[t]
	\includegraphics[width=2\columnwidth]{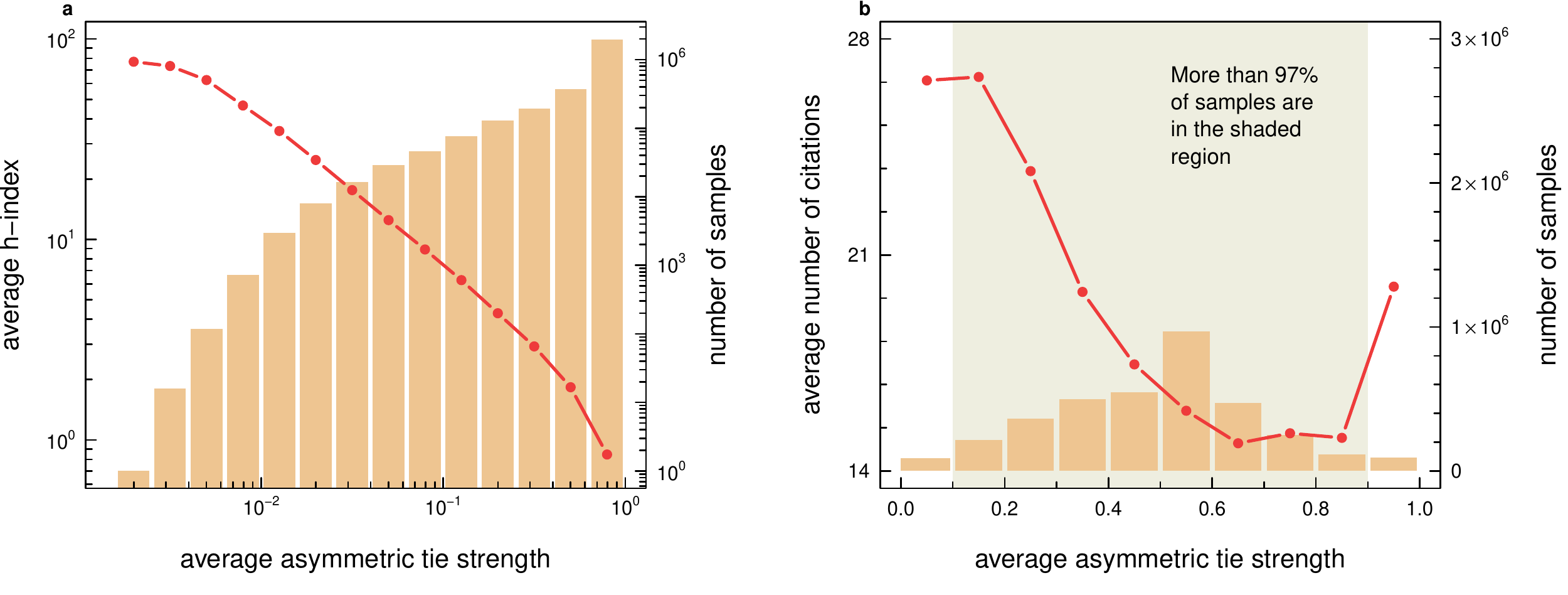}
	\caption{\label{full2hipothesis}\textbf{The role of tie configuration on scientific success of researchers and publications}. \textbf{a} -- the mean h-index of scientists characterized by a given average asymmetric tie strength $\langle v_i\rangle$, Eq.~(\ref{meanvij}). \textbf{b} -- the average number of citations obtained by papers created in teams with a given average asymmetric tie strength. The decreasing nature of both empirical relationships (\textbf{a}~and~\textbf{b}) clearly indicates that scientific collaboration based on weak ties is more appreciated in terms of the number of citations. Moreover, since the number of citations is often correlated with the quality of research, the above results also show that weak ties usually result in more creative (in terms of knowledge production) scientific collaborations.}
\end{figure*}

Now, using the concept of asymmetric tie strength, we will discuss  Granovetter's second hypothesis, which postulates that although weak ties do not carry as much communication as strong ties do, they often act as bridges, providing novel, non-redundant information, which guarantees weakly connected nodes generally understood social success. 

In scientific collaboration networks, the validity of Granovetter's second hypothesis has never been tested. Nevertheless, it is widely believed (see \cite{2016ResPolWang} and references therein) that information and expertise at the disposal of tightly connected research groups are often redundant, resulting in less creative collaborations and less innovative publications, while intergroup collaborations that bridge the so-called \textit{structural holes} \cite{1992bookBurt, 2004JAmSocBurt, 2007JEconTheoryGoyal} can provide access to information and resources beyond those available in densely connected communities, thus leading to novel ideas and valuable publications. To quantitatively address these issues, we check whether the bibliometric indexes of scientists and publications are correlated with the tie strength of the scientific collaboration network. Specifically, we focus on two questions: i) How does the researcher's h-index depend on the structure of his/her local collaboration network? ii) How does the strength of the ties between scientists influence the success of their joint publication? 

\begin{figure*}[t]
	\includegraphics[width=2\columnwidth]{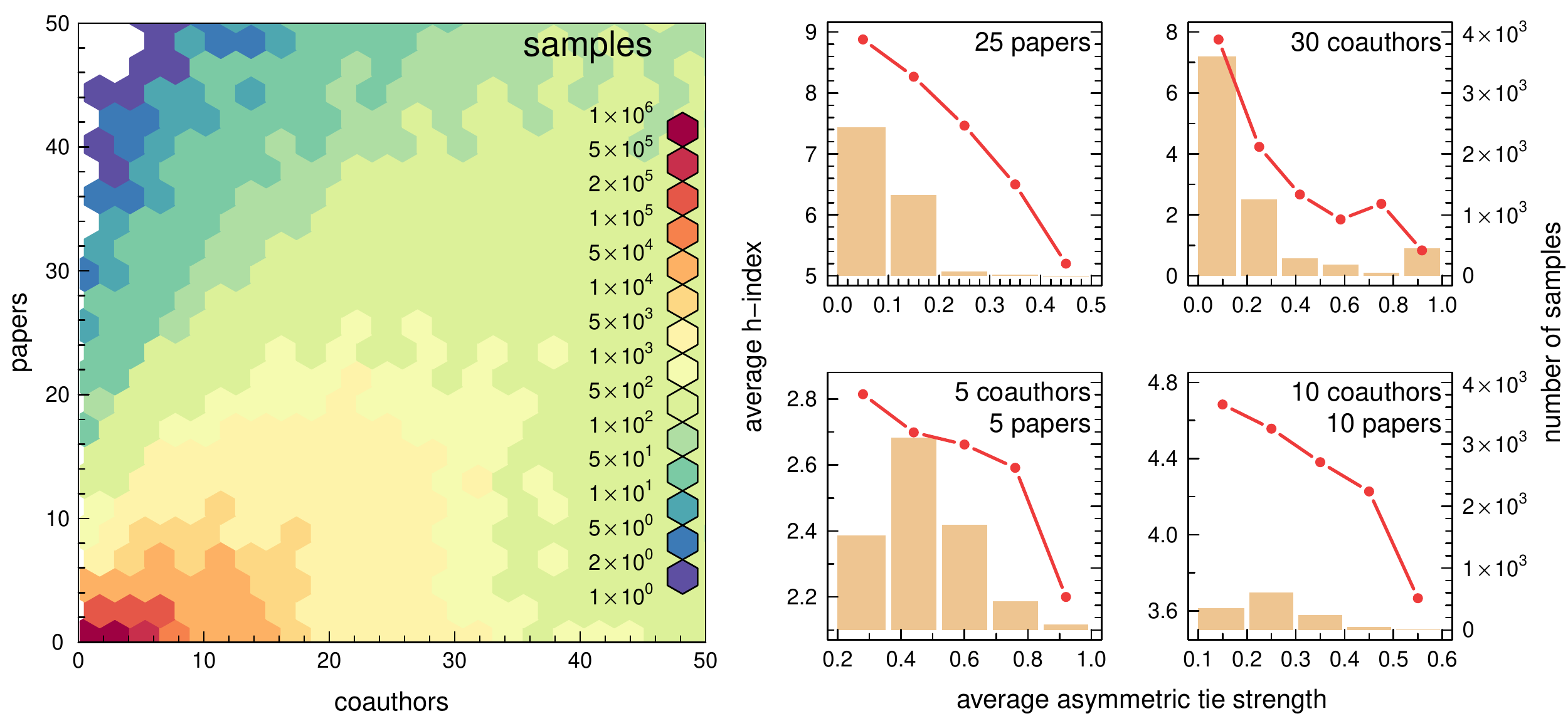}
	\caption{\label{panelhindex}\textbf{Scientists' h-index vs. tie strength}. In this figure, we present a more detailed analysis of the relationship from Fig.~\ref{full2hipothesis}a, which shows the data averaged over all scientists in the studied collaboration network, regardless of the stage of their scientific career. Here we divide scientists into groups in which everyone has the same number of total publications and the same number of co-authors (see the colour map in the figure). The more homogeneous conditions thus established allow us to clearly confirm earlier findings. In particular, as one can see in the small graphs on the right side of the colour map, regardless of the choice of the homogeneous group of scientists their h-index always decreases with increasing average asymmetric tie strength.}
\end{figure*} 

\begin{figure*}[t]
	\includegraphics[width=2\columnwidth]{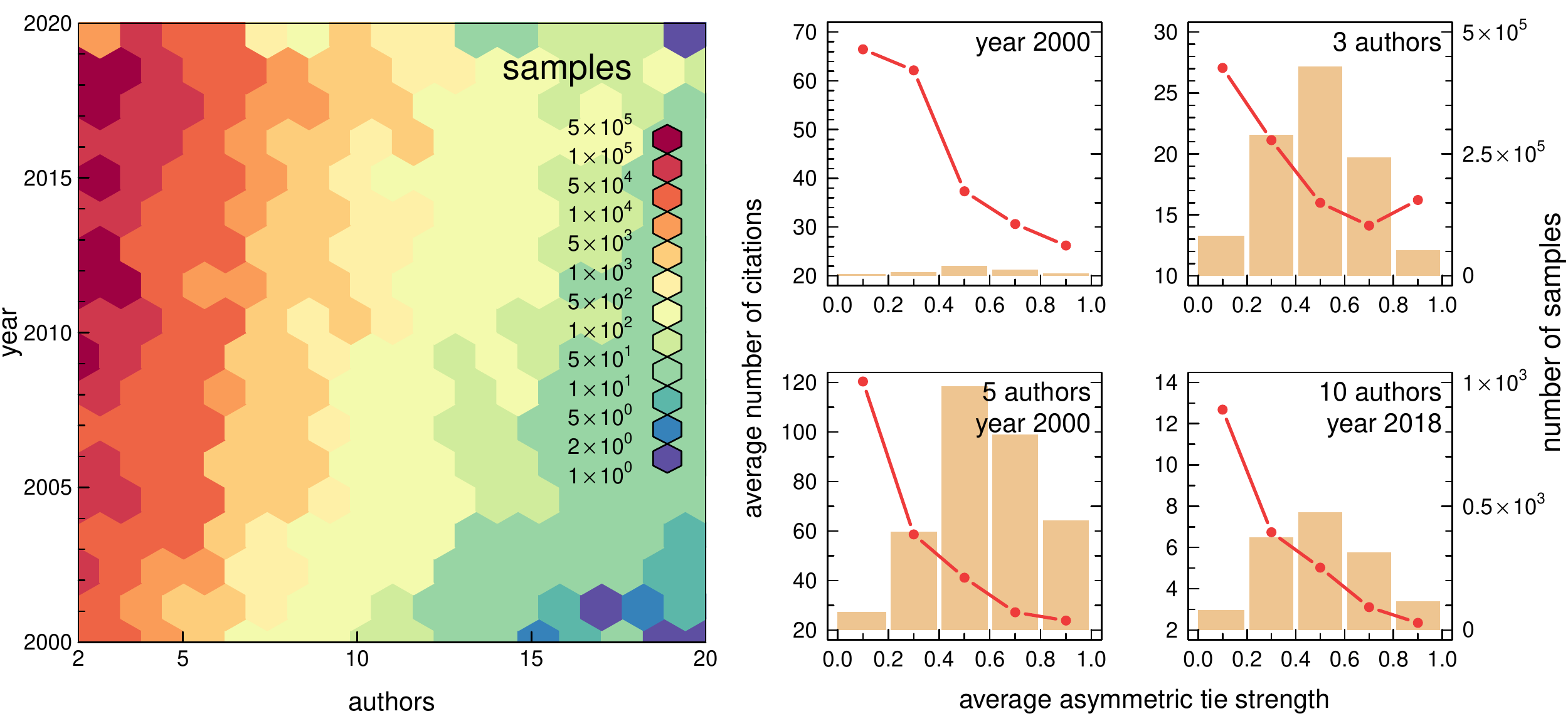}
	\caption{\label{panelcitations}\textbf{Citations of publications vs. tie strength}. In this figure, we present a more detailed analysis of the relationship from Fig.~\ref{full2hipothesis}b. To this aim, all publications available in the analysed database are divided into groups according to the year of publication and the number of authors (see the colour map in the figure). Given homogeneous sample of publications thus established, we found that the number of their citations always decreases with increasing average asymmetric tie strength between their authors. To clarify, the average tie strength was determined at the time of paper's publication, and the number of citations refers to the time of the last update of the analysed database.}
\end{figure*}

To answer the first question, we examined how the h-index  \cite{2005PNASHirsch, 2015NatPhysDorogovtsev} of a scientist depends on his or her average asymmetric tie strength  (see Fig.~\ref{meanv}):
\begin{equation}\label{meanvij}
	\langle v_i\rangle=\frac{1}{k_i}\sum_{j}v_{ij}.
\end{equation}
Eq. (4) quantitatively measures the tendency of scientists to keep collaborating with the same people (cf. the concept of \textit{social inertia} \cite{2006PRERamasco, 2007EPJRamasco}). Fig.~\ref{full2hipothesis}a shows that the averaged (over all scientists who have a similar average tie strength) h-index decreases with $\langle v_i\rangle$. It means that successful (double-digit h-index) scientists have significantly weaker ties than less successful (single-digit h-index) researchers. The result is consistent with Granovetter's general understanding of the role of weak and strong ties. However, since some doubts may arise from the fact that the data presented in Fig.~\ref{full2hipothesis}a are averaged over many different scientists (having a small and large number of all publications, with a small and very extensive network of collaborators), in Fig.~\ref{panelhindex}, we demonstrate that the decreasing nature of the relationship between the h-index and tie strength is independent of the choice of a group of scientists. That is, it still decrease, even in very homogeneous (in terms of scientific achievements) groups of researchers. In particular, as one can see in the small graphs accompanying the colour histogram that represents the available scientists' samples, of any two researchers who have the same number of publications and/or co-authors, the one with weaker ties tends to have the higher h-index. In a way, this suggests that being a good manager and skilfully planning one's network of scientific contacts ensures success \cite{2015PNASPetersen}. This conclusion, however alarming as it may seem, finds its basis in the theory of social networks - the already mentioned concept of Burt's structural holes and social capital \cite{1992bookBurt, 2004JAmSocBurt}.

The role of weak ties in scientific success is even more apparent in relation to scientific publications. Fig.~\ref{full2hipothesis}b shows how the number of citations of a scientific paper depends on the asymmetric tie strength (averaged over all co-authors of each article). The decreasing nature of this relationship indicates that publications created by teams of scientists linked by weak ties are better cited than those that arise in teams with strong ties. In Fig.~\ref{panelcitations}, by analysing more homogeneous samples of publications (published in the same year and/or by the same number of co-authors), we clearly confirm the validity of the above finding. Furthermore, although the number of citations does not always translate into the quality of the research presented, it is undoubtedly a measure of the commercial success of a publication and a specific measure of the knowledge diffusion in scientific collaboration networks.

\section*{Data description and availability} 
The research presented in this paper is based on the publicly and freely available Citation Network Dataset \cite{2008SIGKDDTang}. We used the 12th version of the dataset (DBLP-Citation-network V12) which contains detailed information  (i.e., year of publication, journal, number of citations, references, list of authors) and approximately 5 million articles published mostly during the last 20 years.

It is important to note that our analysis is limited to the largest connected component (LCC) in the co-authorship network, which can be recreated using the dataset. LCC comprises of close to three million nodes (authors), which means it spans 65\% of the entire network. These nodes are connected by more than 13 million bi-directional co-authorship edges.

While the dataset provides exhaustive information about published papers, it does not directly contain any bibliometric information about authors. However, it is possible to calculate various bibliometric indicators either by recreating the network of citations or by directly using article metadata available in the dataset for each article (such as the number of citations). In order to calculate the h-index for all authors in the LCC, we decided to rely on the latter method and use article metadata to determine the number of citations. Considering that the citation network recreated from the dataset is only a sample of the full citation network, this method is more reliable. The number of citations calculated by counting links in the citation network is, in general, underestimated when compared with the number of citations available in the article's metadata.


\section*{Code availability} 
The code that supports the findings of this study is available from the corresponding author upon reasonable request.


\end{document}